# Estimating Multiple Step Shifts in a Gaussian Process Mean with an Application to Phase I Control Chart Analysis

Issac Shams, Saeede Ajorlou, and Kai Yang

*Abstract* − **In preliminary analysis of control charts, one may encounter multiple shifts and/or outliers especially with a large number of observations. The following paper addresses this problem. A statistical model for detecting and estimating multiple change points in a finite batch of retrospective (phase I) data is proposed based on Likelihood Ratio Test. We consider a univariate normal distribution with multiple step shifts occurred in predefined locations of process mean. A numerical example is performed to illustrate the efficiency of our method. Finally, Performance comparisons, based on accuracy measures and precision measures, are explored through simulation studies.**

## I. INTRODUCTION

Statistical Process Control (SPC) was introduced by Walter A. Shewhart in the 1920s as an attempt to present a powerful collection of statistical and also managerial techniques to monitor product quality and maintain process stability through the reduction of variability. Of all the SPC tools, Control Charts are the most popular on-line procedures assisting engineers to quickly detect the happening of assignable causes of process shifts by signaling out-of-control (OC) alarms [1]. Most of control charting methods and corresponding diagnostic tools deal, directly or indirectly, with prospective applications, also called phase II applications, where true in-control (IC) process parameters are accurately estimated or assumed to be known. In this stage, as each new reading (from one or more quality characteristics) obtains successively, the sample statistic is calculated and the SPC check, whether an OC condition has occurred or not, is re-applied [2]. However, every process monitoring has an early stage, namely phase I, in which a finite set of historical data is collected, when the process is thought to be IC, and analyzed all at once as a batch. In the simplest case, data on *m* subgroups are assumed to be independent and identically distributed (i.i.d) random variables, which follow a known density function. On the other hand, there are many practical cases in which these assumptions are not simply satisfied.

The main purpose of phase I analysis, also called retrospective or stage 1 analysis, is to simultaneously detect any special causes of variation, eliminate root causes of these problems, bring the process back to a state of statistical control, and estimate the IC process parameters [3]. In such actions, one may confront multiple shifts and/or outliers (single excursions resulted from momentary assignable causes) especially within a large number of observations.

The general statement for change point problem is as follows. Suppose that statistical data are obtained as an output of a specific experiment, or are being accumulated during such experiment. In the first situation, the entire array of data (with fixed length) is analyzed with the purpose of building a mathematical model or estimating some parameters of interest. As a rule, prior to parameter estimation and model generation, one has to check the hypothesis of homogeneity of data acquired. If this hypothesis is rejected, then segments of homogeneity of data should be detected and parameter estimation should be performed in each segment separately. This is due to the fact that one cannot statistically estimate those parameters that have changed in data acquisition process. In the second case, data are received sequentially in such a way that forms an endless stream. So, any disturbance in the stochastic homogeneity of data being obtained might be an indication of specific event (failure, malfunction, etc.) and should be dealt with on-line to avoid possible losses and casualties [4]. The point at which disruption has been plugged into the data is called change point. In SPC framework the former case is related to preliminary applications whereas the latter pertained to phase II problems.

There are some main classifications of change point problems found in the literature: (1) whether the type of change (step, trend, monotonic, sporadic) is known as a *priori* or not; (2) whether the number of changes (single or multiple) is known exactly a *priori* or not; (3) whether the assumption of independence of observations is contradicted or not; (4) the different volume of priori statistical information about underlying process which leads to parametric, semiparametric or nonparametric methods of change point detection, etc.. Studies carried out by [5, 6], and [7] are the initial attempts in the literature dealt with a *posteriori* change-point problem with one abrupt change at the unknown moment of density function, and sequential change-point detection respectively. [8], then proposed Bayesian and minimax procedures in order to solve the problem of optimal sequential change point detection. After that, the change point analysis is intensively investigated by [9], [10, 11], [12], and others and has been successfully applied in different fields including statistical process monitoring, statistical control theory, pattern recognition, signal processing, etc. In SPC context, most studies of change-point problem devoted to ongoing (phase II) settings where the emphasis is on process monitoring with a clean set data obtained from retrospective analysis. In this application, the process is first assumed to be in-control and monitored by a specific control chart. Then a special cause,

Issac Shams is with the Department of Industrial and Systems Engineering, Wayne State University, Detroit, MI, 48202 USA. (e-mail: er7671@wayne.edu).
Saeede Ajorlou, is with the Department of Industrial and Systems Engineering, Wayne State University, Detroit, MI, 48202 USA. (e-mail: er7212@wayne.edu).
Kai Yang, is with the Department of Industrial and Systems Engineering, Wayne State University, Detroit, MI, 48202 USA. (e-mail: ac4505@wayne.edu).

mainly *sustained* special cause, occurs at an unknown point in time, and later an OC condition is usually signaled by the chart. Afterward, the process is, usually, stopped and search for finding a change point is performed. Hence, the problem of multiple change points is rather meaningless in prospective applications (see e.g., [13], [14], [15], [16], and [17]). On the other hand, there exist a few papers considering change point(s) problem in phase I analysis of SPC. In this situation, [3] addresses multiple shifts and/or outliers with a rational subgroup of size one using clustering approach and shows that neither X-chart nor CUSUM chart can detect the existing of any disruptions when multiple shifts and/or outliers are present. In additions, [18] propose a dynamic programming model to find the exact change point locations in a hierarchical clustering approach, subject to the restriction that all segments have at least a specific observations.

Due to the nature of process we study, there are many situations in which it is advisable to work with individual observations; for example the production rate is low or there is an automatic measurement system and every unit can be examined separately (see [1] for more details). The main purpose of this study is to address multiple change point problem in phase I application with individual observations. We propose a method based on Likelihood Ratio Test (LRT) which is able to detect multiple step shifts in a batch of i.i.d normal random variables and also can estimate the correct location of change points. The strong and weak aspects of the method are further discussed in detail.

The remaining of this paper is organized in the following order: In the next section, we elaborately clarify the problem and provide a basis for applying LRT approach to multiple change point's problem with individual observations. In section 3, a comprehensive numerical example is presented to illustrate the validity of our method. Also, performances of the proposed method are assessed by simulation experiments based on accuracy measure and precision measure. Finally, in section 4, we present some conclusions and further research.

II. METHODOLOGY

There are some recommendations proposed for the problem of phase I analysis with individual measurements, most common of which is to use X and MR control chart ([1] or [18]) or just constructing X chart ([19]). However, in this paper we use likelihood ratio test approach to address the problem. In this procedure, all potential segments of the historical (phase I) data set into two subgroups are considered and LRT statistics related to each segment are formed. When one or more computed LRT statistics exceeds a threshold value, an OC condition is indicated. Moreover, the segment corresponding to the maximum value of the statistic is specified as the most likely location of the change. As noted, the LRT method can be applied either to detect the change point or its location in a batch of random input variables. However, in order to use this method, one must first determine the appropriate probability distribution of underlying process.

Suppose a batch of $m$ historical independent observations, $x_1, x_2, \ldots, x_m$, from one or more univariate normal distributions all with same variance $\sigma^2$. There are $R$ shifts in the mean, and the shift locations are $\tau_r$, r=1,…, $R$ subject to $0 < \tau_1 < \ldots < \tau_R < m$ where $\tau_0=0$ and $\tau_{R+1}=m$. Let $\zeta_t(0)$ represents the probability distribution function of $x_t$. So the model can be formulated as

$$x_t : \zeta_r(0) \text{ for } \tau_{r-1} < t \leq \tau_r; \text{ r=1,K },R+1 \tag{1}$$

We want to determine if the process from which the batch is obtained indicates an IC condition, which corresponds to $R=0$; and if not where the exact change point(s) are.

Assume that the first change point occurred in the mean of independent Gaussian variables is located in the $m_1^{\text{th}}$ observation such that $m_1 < m$ and $m_1 + m_2 = m$. The log of the likelihood function for the first $m_1$ observations is

$$-\frac{m_1}{2}\ln(2\pi\sigma^2) - m_1\frac{\hat{\sigma}_1^2}{2\sigma^2} - m_1\frac{(\bar{x}_1 - \mu)^2}{2\sigma^2} \tag{2}$$

This term is maximized for

$$\bar{x}_1 = \frac{1}{m_1}\sum_{t}^{m_1} x_t \text{ and}$$

$$\hat{\sigma}_1^2 = \frac{1}{m_1}\sum_{t}^{m_1}(x_t - \bar{x}_1)^2 \tag{3}$$

which are the Maximum Likelihood Estimators (MLE) for the first $m_1$ observations. The maximized value for (2) is

$$l_1 = -\frac{m_1}{2}\ln(2\pi) - \frac{m_1}{2}\ln(\hat{\sigma}_1^2) - \frac{m_1}{2}. \tag{4}$$

Similarly, the maximized value of likelihood function for the remaining $m_2$ observations is

$$l_2 = -\frac{m_2}{2}\ln(2\pi) - \frac{m_1}{2}\ln(\hat{\sigma}_2^2) - \frac{m_2}{2}. \tag{5}$$

Therefore, in the case that there are multiple change points within a preliminary data set, the maximum log-likelihood function for all observations is given as follows:

$$l_a = l_1 + l_2 \tag{6}$$

On the other hand, if the process was IC then all $m$ observations are identically distributed and the maximized value of likelihood function can be found as follows

$$l_0 = -\frac{m}{2}\ln(2\pi) - \frac{m}{2}\ln(\hat{\sigma}^2) - \frac{m}{2}. \tag{7}$$

In case that $l_a$ is substantially larger than $l_0$ the process is considered to be OC. Minus two times the difference of the log-likelihood function

$$\begin{aligned}lrt[m_1, m_2] &= -2(l_0 - l_a) \\ &= m\ln(\hat{\sigma}^2) - m_1\ln(\hat{\sigma}_1^2) - m_2\ln(\hat{\sigma}_2^2)\end{aligned} \tag{8}$$

has asymptotically chi square distribution with two degrees of freedom. For large sample approximation see [20].

In summary, the LRT method computes (8) for all possible values of $m_1$ and introduces one as the mle for the change location which maximizes the (8), given that the maximized value exceeds a predefined threshold value. Clearly, a control chart can be set up by plotting the statistic (8) versus $m_1$ signaling an OC condition if any value goes beyond an upper control limit. Moreover, the method can be applied to detect multiple shifts, especially with a large number of observations, by binary segmentation. If a change is detected, then the data will be divided at the most likely location for a single change, and the procedure is repeated to each new group. This continues until no subgroup shows evidence of any change.

Recall that the stage 1 situation for individual NID random variables with known $\sigma^2$ is the focus of this study. However, the true variance of a batch of historical observations is not known for almost all real cases and is difficult to be accurately estimated with individual observations. This problem becomes worse when multiple shifts and/or outliers are present. So it is of interest to build a robust estimator in this situation and compare it with typical ones based on the average of the moving ranges or sample standard deviation. Furthermore, one can consider multiple shifts in both mean and variance, separately or simultaneously, within a batch of retrospective data and customize LRT approach to be able to attribute a signal to a shift in mean only, variance only, or a combination.

In the next section, we statistically compare the efficacy of our proposed methods using Monte Carlo simulation.

III. NUMERICAL EXAMPLE

In this section, we employ Monte Carlo simulation to effectively study the performance of the proposed method. Two commonly-used measures, accuracy and precision measures are provided to evaluate the efficiency of change point estimators. The former sizes how close an estimated value is to the real value whereas the later rates how close the estimated values are to each other.

It is assumed that there are $R$ step change(s) with different magnitudes of $\delta$= 0.5, 1.0, 2.0, 3.0, 4.0 and 5.0 in the mean of a batch of $m$ univariate normal random variables. On the grounds that it is desirable to have 20-25 samples of size 3-5 for constructing a trial control chart of preliminary data [1], we set the value of $m$ to be 200 in our study.

Moreover, as stated in [1], there are various types of pattern which may be revealed in a specific control chart, fairly often in phase I, as results of some particular inputs. Interpretation of such patterns provides valuable diagnostic information on the process and also helps bring a process back to IC condition for prospective applications. In this study we assume that a Mixture pattern, which best serves our purpose, is fed into the batch as a result of two overlapping distributions generating the process output (see [1] for more details about common causes and effects of a Mixture pattern on control charts).

Also, we consider, for multiple change point situation, uniformly spaced shifts alternating between two means. That is, a single shift occurs midway in the data, and two shifts would be located after one-third and two-thirds of the observations and so on. One may take into account random shift locations [22,23], but there is a potential risk in this condition; if the shifts position near the end of the batch or close to another shift, they may resemble outliers. To this end, presume $\Omega = \{\mu_{j+1} \mid \mu_{j+1} = \mu_j + \delta * (-1)^j ; j = 0,1,\ldots,R-1\}$ is the set of change values, $\mu_0$ is predetermined initial value, and $\delta$ is the magnitude of change. Without any loss of generality $\sigma = 1$ is used in the simulation. $\Omega$ is defined such that the difference in parameter mean for two consecutive groups is identical and equal to $\delta$. For example, imagine that there are $R$=4 groups with different mean values and let $\mu_0 = 1$ and $\delta = 3$. In this case, sequence $\Omega$ is defined as $\Omega = \{1, 4, 1, 4\}$. So, there are five different groups: the first group consisting of the first observation to the 40$^{th}$ observation, all are randomly generated from a NID (1, 1), the second group consisting of the 41$^{th}$ observation to the 80$^{th}$ observation, all are randomly generated from a NID (4, 1), and so on.

Owing to the fact that the expected value of the statistic (8) is not consistent varying the value of $m_1$ (in fact if $m_1$ or $m_2$ is small, the expected value is always larger than when both are the same), as recommended by [21], it is desirable to improve the test by diving each test statistic by its in-control expected value. Then the threshold value(s) is determined using simulation to give the desired false-alarm probability [24,25]. Therefore, normalizing the statistic by its expected value gives the new test statistic

$$Nlrt(m_1, m_2) = \frac{lrt(m_1, m_2)}{E\left[lrt(m_1, m_2)\right]} \qquad (9)$$

In this way the resulting expected value is the same for all values of $m_1$. So we apply the improved test statistic (9) instead of (8) in our study.

In order to estimate $\tau_r$'s, 1000 replications are used in each simulation run. Here, we also design our method to be able to perform seven hypothesis tests and detect and estimate maximum seven change points. An alternative model could be built to detect and estimate more or less shifts.

The proposed method is as follows. Using 5,000 independent simulations, we first calculate $E[lrt(m_1,m_2)]$ array, when process is IC, for all possible values of $m_1$ while $m$=200. Then we utilize statistic (9) and start generating a batch of historical data following change set $\Omega$. At each run, seven potential test are conducted and at most $2^0+2^1+2^2=7$ partitions are introduced as the estimates ($\hat{\tau}_r$) of true change points. Also, using 5000 simulations, the false alarm probabilities of 0.03, 0.02, 0.02, 0.01, 0.01, 0.01, 0.01, and 0.01 correspond to threshold values 7.0089, 7.5745, 7.3684, 8.3876, 8.1206, 8.0292, and 7.9153 respectively.

A. *Performance Comparisons Based on Accuracy Measures*

Table 1 shows the estimates of change points and their related standard error (in the parentheses) where there are $R$=1, 2, 3, and 4 step shifts, with different shift sizes, in a batch of $m$=200 retrospective data. It is observed that the proposed LRT method works appropriately providing approximately unbiased estimates for true change locations in case of single change. For other cases, although there is a small biasness, particularly in intermediate change points, the magnitude of biasness is not very large to seriously affect groups. It is worth mentioning that for $R=3$ the method tends to underestimate midway shift and overestimate two other shifts. This tendency exists for $R=4$ in a different manner; most of the time the first two changes are underestimated and the last ones are overestimated. For instance, when $R=3$ and $\delta=2$, the estimates for true change points 50, 100, and 150 are 51.3, 98.7, and 151.6 respectively. Again when $R=4$ and $\delta=3$, the estimations for real change points 40, 80, 120, and 160 are 39.3, 79.2, 122, and 160.3 respectively. In Contrary, LRT method does always overestimate the true change locations for $R=1$ and $R=2$. In additions, the results of table 1 indicate, for almost every case, a progressive increase in accuracy of estimated change points with increment in shift size.

### B. Performance Comparisons Based on Precision Measures

Regardless of the fact that the average of change points can be applied as a summarized comparison among estimators, to explicitly investigate the performances, the proximity of estimates to each other should be taken into consideration. An estimator with good performances in estimating location of changes may inherently have poor performances in terms of dispersion. In this situation, the estimator provides estimates that are close to the true location in average but far from each other.

To pursue this goal, we construct confidence intervals for change points and their probabilities with different coverages from 0 to 25. The related results for the proposed method over a range of $\delta$ are illustrated in Table 2. It is shown that the method has totally acceptable precision performances even for small values of $\delta$. For example, when there is a change of size 0.5 in the normal mean 31% of all estimates of $\tau_1$ show the true change value and half of all estimations are two units or less far from the real shift. As expected, the estimated probabilities of confidence intervals increase, for most of the time, as the size of change point increases. Besides, the method can guarantee to identify a location equal to or less than 15 units from the true location approximately in 90% of times for all possible change sizes.

Finally we should point out although the proposed method performs appropriately subjected to multiple changes and is indeed superior to other conventional methods in terms of accuracy and precision measures; it is restricted by distributional assumptions. In other words, knowing the exact distribution of phase I data is the preliminary step in forming the LRT method, which rarely the case for real world problems. Besides, there are some situations in which the two heterogeneous input data do not follow the identical distribution. These obstacles can be modified by using nonparametric methods such as clustering method or by applying Generalized Likelihood Ratio Test (GLRT) for general distributions.

### IV. CONCLUSION

In this paper, based on likelihood ratio test, we propose a model to address multiple step shifts in a mean of independent normal random variables obtained from phase I analysis of SPC. It is shown that the method functions suitably in terms of accuracy measure and precision measure over various ranges of shift sizes. Our approach can be generalized to include more general distributions, i.e. exponential family or normal family distributions, to detect more shift type, i.e. linear trend or sporadic change, and to simultaneously detect shifts in more than one moment of density function, which are planned for future studies. In spite of superiority of LRT method, it should be noted that this method requires the knowledge about the exact distribution of historical data set. This assumption makes the method restricted and degrades its practicability. Besides, in some cases, this data set may follow different distributions in which the observations between groups follow distributions with various functional forms. Thus, the LRT method should develop to a more generalized form that is flexible for such changes. For example, one can derive the LRT statistic of a general family of distributions such as Johnson family distributions or exponential family distributions that can conveniently fit with separate data sets.